%% file: main.tex
\documentclass[11pt]{article}

\usepackage{jinstpub}
\usepackage{url}
\usepackage{hyperref}
\usepackage{upgreek}

\usepackage[toc,page]{appendix}
\usepackage{enumitem}

\newcommand{\st}{\sec\theta}
\newcommand{\smut}{S^\mu}

\newcommand{\smup}{\widehat{S^\mu}}
\newcommand\mystrut{\rule{0pt}{7pt}}

\newcommand{\smuti}{{S\mystrut^\mu}_i}
\newcommand{\smuptemp}{\widehat{S\mystrut^\mu}_i}
\newcommand{\rtm}{t\mystrut^\mu_{1/2}}
\newcommand{\rtmp}{\widehat{t\mystrut^\mu_{1/2}}}

\usepackage{natbib}

\usepackage{xspace}
\def\Offline{\mbox{$\overline{\textrm{Off}}$\hspace{.05em}\protect\raisebox{.4ex}{$\protect\underline{\textrm{line}}$}}\xspace}

\title{Extraction of the Muon Signals Recorded with the Surface Detector of the Pierre Auger
  Observatory Using Recurrent Neural Networks}

\date{}
\emailAdd{auger\_spokespersons@fnal.gov}

\begin{document}

\abstract{The Pierre Auger Observatory, at present the largest cosmic-ray observatory ever
  built, is instrumented with a ground array of 1600 water-Cherenkov detectors,
  known as the Surface Detector (SD). The SD samples the secondary particle
  content (mostly photons, electrons, positrons and muons) of extensive
  air showers initiated by cosmic rays with energies ranging from $10^{17}~$eV 
  up to more than $10^{20}~$eV. Measuring the independent contribution of the muon component to
  the total registered signal is crucial to enhance the capability of the 
  Observatory to estimate the mass of the cosmic rays on an event-by-event basis. 
  However, with the current design of the SD, it is difficult to straightforwardly 
  separate the contributions of muons to the SD time traces from those of photons, electrons and positrons. In this paper, we present a
  method aimed at extracting the muon component of the time traces
  registered with each individual detector of the SD using Recurrent Neural Networks.
  We derive the performances of the method by training the neural network on 
  simulations, in which the muon and the electromagnetic components of the traces 
  are known. We conclude this work showing the performance of this method on 
  experimental data of the Pierre Auger Observatory. We find that our predictions agree with the parameterizations obtained by the AGASA collaboration to describe the lateral distributions of the electromagnetic and muonic components of extensive air showers.}


\arxivnumber{2103.11983}

\author{\includegraphics[height=30mm]{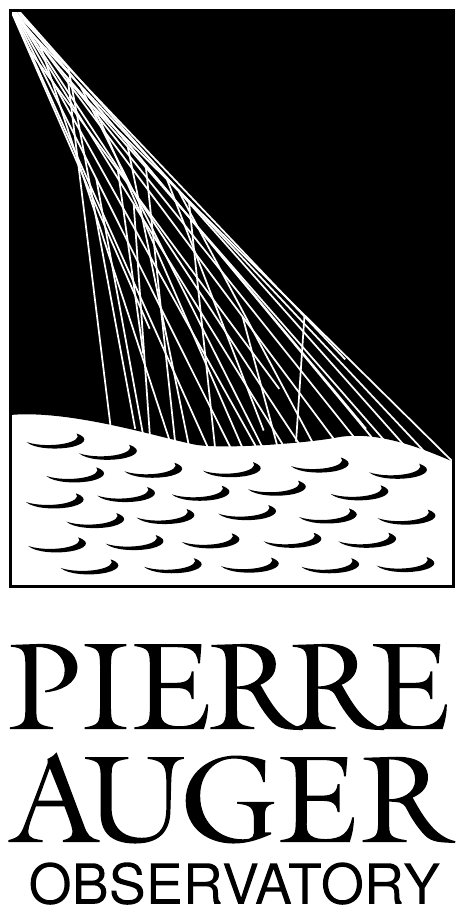}\\[3mm]The Pierre Auger Collaboration}
\affiliation{Av.\ San Mart\'{\i}n Norte 306, 5613 Malarg\"ue, Mendoza, Argentina}

\maketitle

\section{Introduction}
\label{sec:introduction}

The existence of ultra-high energy cosmic rays (UHECRs) is one of the most intriguing phenomena
in astroparticle physics. UHECRs are atomic nuclei with energies above
$10^{18}$\,eV and their flux falls off quickly with energy \cite{Aab:2020rhr,Aab:2020gxe}. 
The most energetic particles, above $10^{20}$\,eV, are detected with a frequency smaller
than one per square kilometer per five centuries. The Pierre Auger Observatory was
designed and built to unveil the origin of these particles~\cite{ThePierreAuger:2015rma}. It is the largest cosmic-ray
detector built so far and comprises an area of 3000\,km$^2$. The detection of
air showers is performed using two complementary techniques. On the one hand, 
the fluorescence light produced as the shower develops along the atmosphere 
is measured at four sites of the Fluorescence Detectors (FD). On the
other hand, the particles that reach the ground are sampled with the SD, an array of 
1600 water-Cherenkov detectors (WCDs), separated 1500\,m from each other.

\begin{figure}[t]
  \centering
  \includegraphics[width=.47\textwidth,]{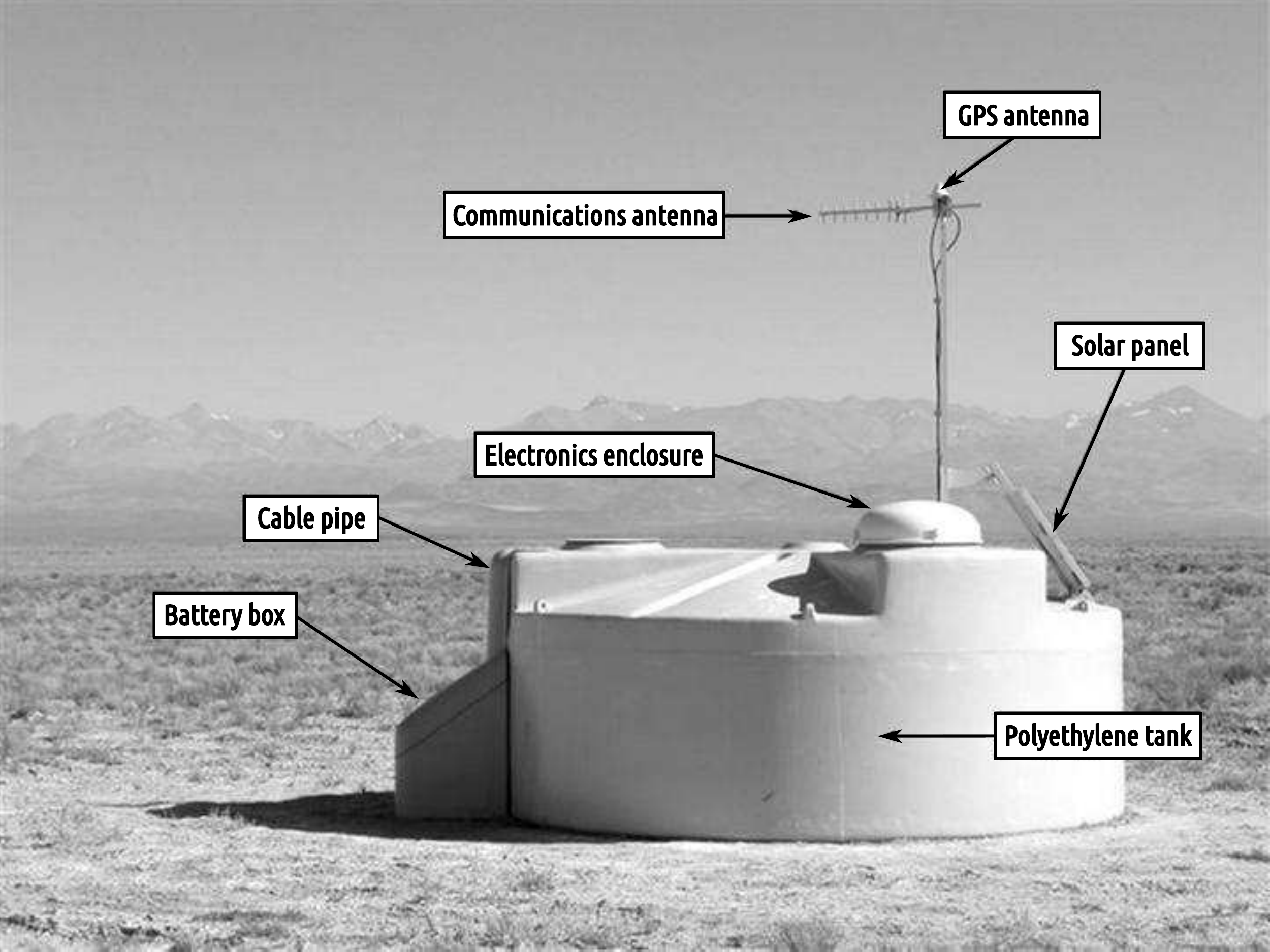}
  \hspace{.1cm}
  \includegraphics[width=.47\textwidth,trim={0 .3cm 0 0},clip]{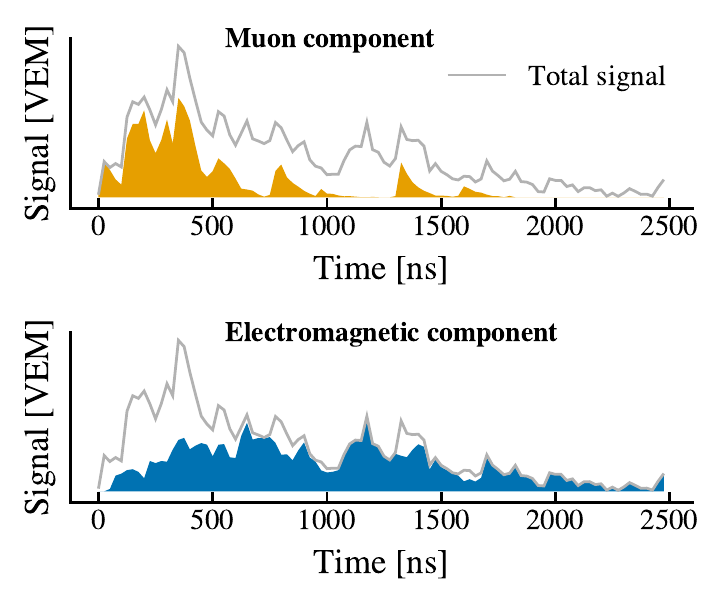}
  \caption{
    Left: A schematic view of a surface detector station in the field, showing
    its main components.
    Right: Signal in time measured by a single
    station of the SD for a simulated event. The total signal corresponds to the
    signal that can be measured by the SD. The signal from the electromagnetic and muon
    components are shown independently, as this information is available in
    simulations. The total signal is the sum of the muon component and the
    electromagnetic component. Our goal is to obtain the muon component using the total signal as an input.}
  \label{fig:sd}
\end{figure}

This work focuses on the data collected with the SD, which operates with a duty cycle close to 100\% and shows
full detection efficiency for air showers above 3.16\,EeV ($10^{18.5}$\,eV). Charged particles traverse the 12\,000 litres of ultra-pure water contained in
each station of the SD and the emitted Cherenkov photons are collected by three nine-inch photomultiplier tubes (PMTs); see the left panel of
\autoref{fig:sd}. The photomultiplier output is digitised at 40 MHz (25 ns bins) using
10 bit Flash Analog-Digital Converters (FADCs). Each signal trace has a length of
768 samples, corresponding to a time window of 19.2 $\upmu$s. The resulting signal is proportional to the sum of the 
Cherenkov radiation in water
produced by electromagnetic
particles (photons, electrons and positrons) and muons, see the right panel of
\autoref{fig:sd}. Muons produce some characteristic features in the signal. As highly penetrating particles, they usually arrive at the stations at earlier times
than the electromagnetic particles that undergo copious multiple scattering, see Fig.~4 in ref.~\cite{1962PhRv..128.2384L}. 
The signal that muons produce is spiky, in
contrast to the signal from electromagnetic particles, which is more continuous and spread in
time. Since muons suffer less energy losses in the atmosphere than electromagnetic
particles~\cite{Gaisser:2016uoy}, they dominate the signals of the stations located far
away from the core of the shower.

Based on the Heitler-Matthews extensive air-shower model and assuming the
superposition principle \cite{Matthews:2005sd}, the number of muons $N_\mu^A$ in
an extensive air shower produced by a nucleus with mass number $A$ can be
related to the number of muons produced in a shower initiated by a proton with
the same energy, $N_\mu^\text{p}$, through
\begin{equation}
N_\mu^A=N_\mu^\text{p} A^{1-\beta}.
\end{equation}
where $1-\beta\simeq 0.1$.
Therefore, the determination of the muon component in the air shower is crucial
to infer, for each event, the mass of the primary particle, which is a key
ingredient in the searches conducted to pinpoint the sources of UHECRs. Mass
composition inferences are customarily carried out comparing observables
measured in data and simulations.

With the current design of the SD, the separation of the muon and 
electromagnetic components can be done for events with large zenith angles \cite{Aab:2014pza} or for distances far off from the shower core \cite{Aab:2014dua}.
Hence, for a majority of the recorded data, this separation cannot be performed
in a straightforward and efficient way.  
In this work, starting from the total time trace, we estimate the
muon signal for each individual WCD as a function of time using machine learning techniques. The
applicability of the method is not restricted to a limited energy and/or zenith
angle range but can be applied to the full data sample
collected so far. However, for practical purposes, in this work we limit the
discussion to the data set referred to as ``vertical'' events
($\theta < 60^{\circ}$, where $\theta$ is the zenith angle) which are those for which it is harder to estimate the muon component with more straightforward techniques. As a novelty, we estimate the risetime \cite{Aab:2017cgk} of the muon signal.

The branch of machine learning research studies algorithms and techniques that rely on
inferring patterns from data. The scientist is only in charge of developing a
suitable architecture, instead of designing every detail of a model. Numerous
techniques were proposed many years ago but several recent breakthroughs in
computational hardware have allowed for these techniques to be computationally
affordable. Nowadays, machine learning is being used extensively in physics (for
a recent review see e.g. ref.~\cite{Carleo:2019ptp}). It is especially well 
suited for experiments in particle physics~\cite{HEPML} and in astroparticle physics. The large amount of events collected by state-of-the-art experiments in 
both fields allows for the exploitation of machine learning techniques that
require vast amounts of data to build and train a model.

An example of a previous work devoted to the extraction of the muon signal using machine
learning can be found in ref.~\cite{Guillen:2018uex}. In this work, we focus our
interest on the study of time series. Hence, from the large set of machine learning
algorithms, we have chosen deep learning since it is the one that has lately
experienced substantial progress when applied to temporal series and sequences.
We go one step beyond to what is discussed in ref.~\cite{Guillen:2018uex}, where
only the integral of the muon trace was obtained. For the first time, we
estimate the muon contribution to the signal recorded in each time bin.

This article is structured as follows. In \autoref{sec:method}, we give an
overview of the methodology, the architecture of the neural network and the training method. In \autoref{sec:results}, we show the results of the neural
network. We conclude by discussing the application of
the neural network to data and assess its performance scrutinizing standard properties of extensive air showers.

\section{The method}
\label{sec:method}

Our approach is based on Recurrent Neural
Networks (RNNs). RNNs are specially well-suited for time series due to their
memory mechanism. As the computing process develops, for each step of the
temporal series, the RNN stores information from a preceding time slot that can
be used at a later stage. Nowadays, RNNs are used successfully in different
fields such as natural language processing or machine translation, see for
example \cite{effectiveness_of_rnn}. There are several kinds of RNNs and, among
them, we have chosen one of the most common, known as Long Short Term Memory
(LSTM) \cite{Hochreiter:1997yld, understanding-lstm}.

\subsection{The input}
The traces recorded by the water-Cherenkov detectors of the SD are the input to the
neural network. The electronics is used to sample the signals in bins of 25 ns. Once
they are calibrated, the signals are expressed in Vertical Equivalent Muons
(VEMs), which correspond to the most-likely signal deposited by a  muon that
traverses the center of the detector top-down. The traces used in this study are obtained after averaging the
traces of every functioning PMT in a WCD. 

We use only the first 200 bins (5\,$\upmu$s) of each trace. Traces that are shorter than those
200 bins are padded with zeros at the end. This choice is made because the most relevant
information is encapsulated in the first 200 bins of each trace, while choosing too many bins increases the computational time and memory needed to train the neural network. From simulations we
know that for $E<10^{19}$\,eV, around 90\% of the stations have the complete muon
signal in the first 200 bins and the remaining stations have more than 99\% of the muon signal
in those 200 bins. For $E>10^{19}$\,eV, around 70\% of the stations have the
complete muon signal in the first 200 bins and the remaining stations have around 99\% of the
muon signal contained in those 200 bins. The fraction of the muon signal outside the 
first 200 bins is then negligible and can be ignored.

\begin{figure}[t]
  \centering
  \includegraphics[width=.7\textwidth]{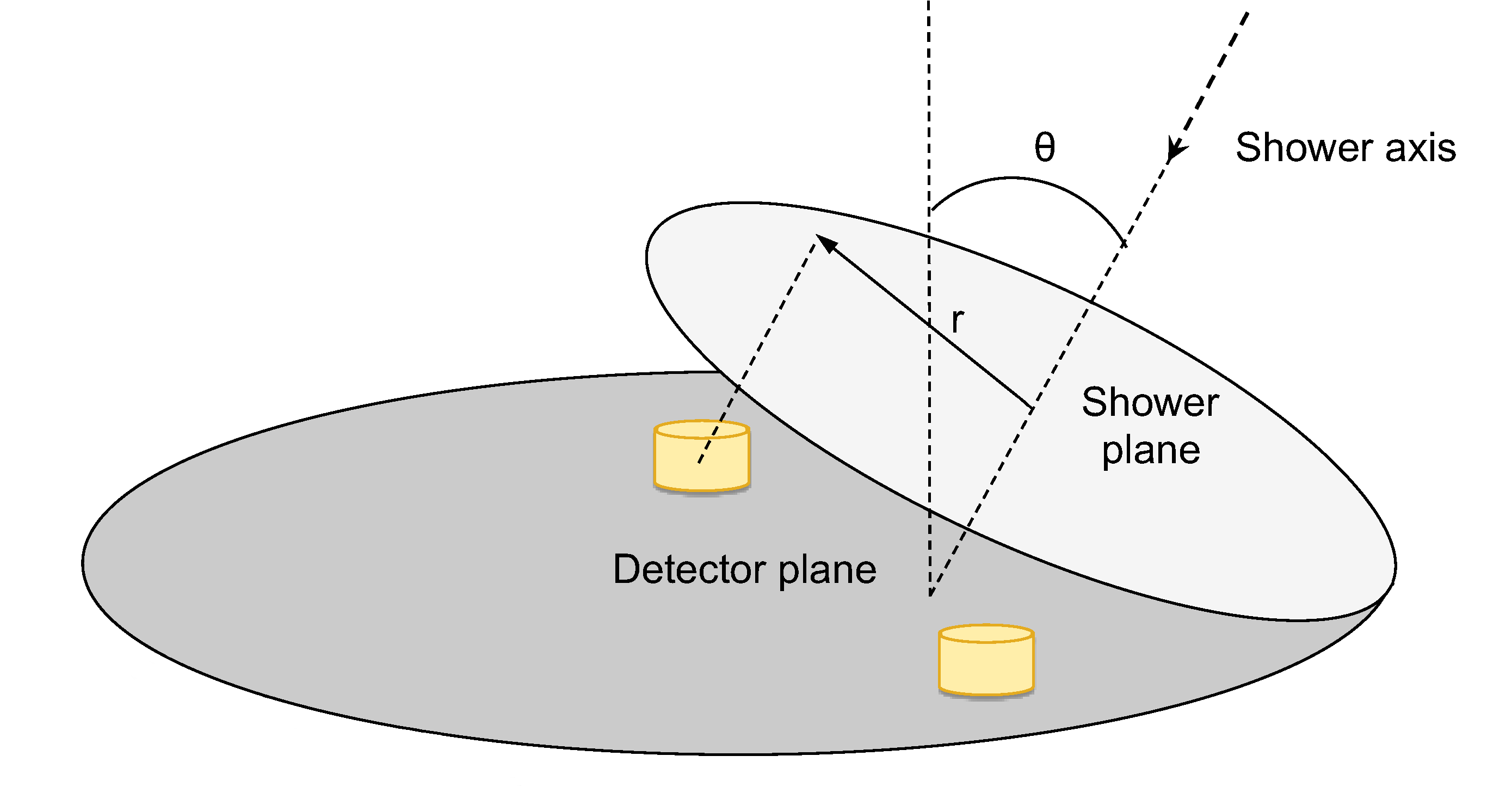}
  \caption{Schematic diagram of the geometry of an event, modified from ref.~\cite{Aab:2016enk}. The angle $\theta$ is the angle between the
    shower axis and the zenith, the distance $r$ is the distance between the
    shower axis and the perpendicular projection of the station in the shower plane.}
  \label{fig:geometry}
\end{figure}

The trace information is not enough to accurately determine the muon component
of extensive air showers. We found that the amount of atmosphere that the
particles traverse plays an important role, since the electromagnetic and muon
components are attenuated differently in the atmosphere. Particles from the
electromagnetic component interact and scatter; the cascade continues down to 
energies below 1~MeV until electrons slow down through ionization without further 
radiating. In contrast, muons are very penetrating particles that traverse the atmosphere
practically unaffected. This is why they also arrive earlier than electromagnetic particles.
Muons are typically minimum ionizing particles. Consequently, as the attenuation
of electrons, positrons and photons increases, the traces richer in muons become
spikier and shorter in time.

This is the reason behind our choice of using as input two more variables: the
secant of the reconstructed zenith angle, $\sec\theta$, and the distance to the
core on the plane perpendicular to the shower (the shower plane), $r$, see
\autoref{fig:geometry}. $\theta$ is the reconstructed angle
between the zenith and the trajectory of the primary cosmic ray. It results from a geometrical reconstruction carried out using the arrival time of
particles at the stations and their respective locations~\cite{Aab:2020lhh}. From this
reconstruction, the position of the core of the shower at the ground is
estimated. 
Both variables take into account the amount of atmosphere
crossed by particles. For $\theta\lesssim 80^\circ$, the 
amount of traversed atmosphere is proportional to $\sec\theta$ and as $r$ increases, the distance travelled by particles through the atmosphere gets larger.

Other variables, such as the energy of the primary cosmic ray, do not help to improve the performance of the neural network when they are included. By
not including it, we also avoid the need to match the energy scale of data with
that of simulations, which proves to be a difficult task. The application of this 
method to data is then straightforward since the zenith angle and distance to the 
core depend only on the geometrical reconstruction of the event.

\subsection{Recurrent Neural Network (RNN) architecture}

\begin{figure}[th]
\includegraphics{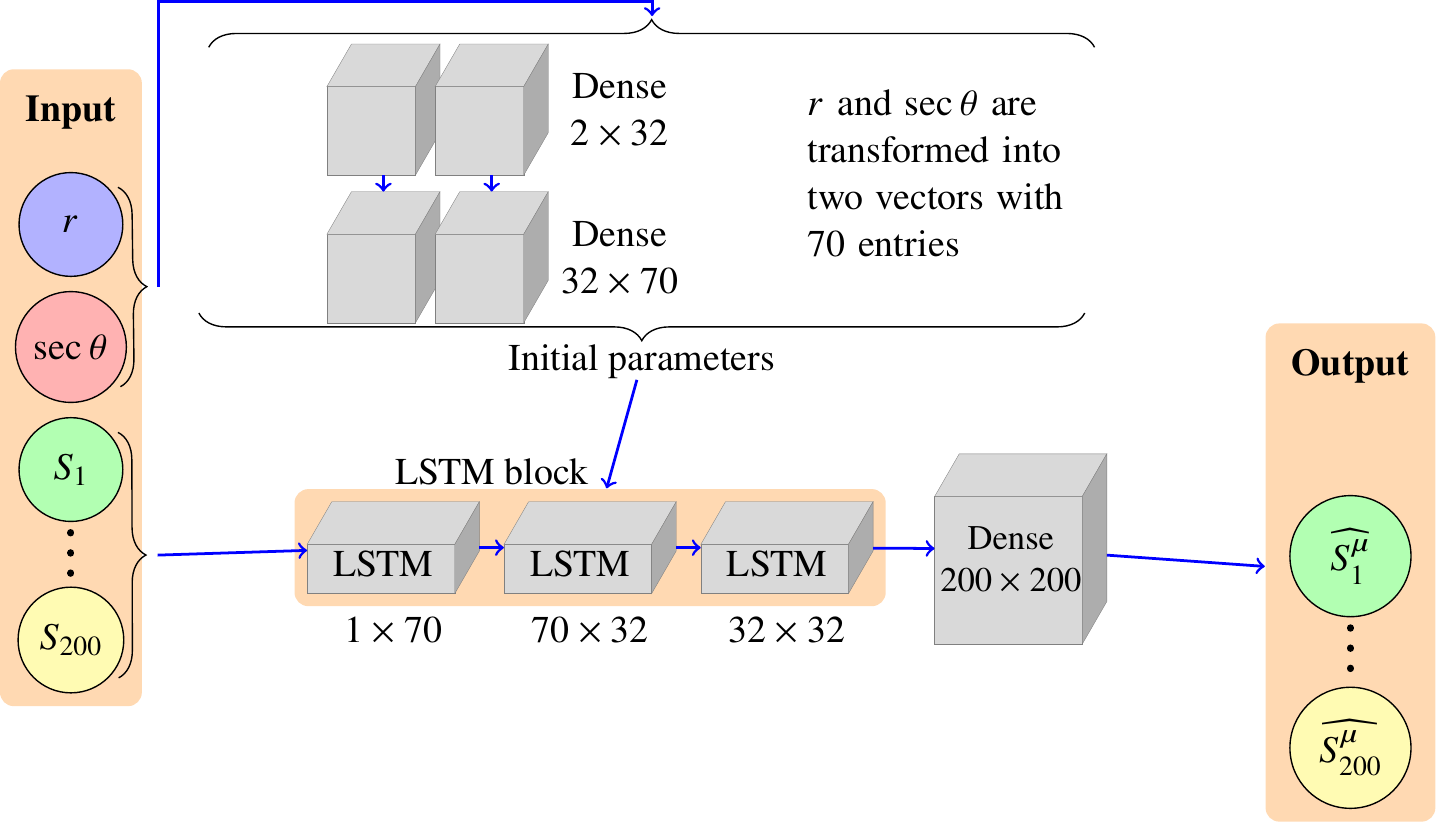}
\caption{Schematic drawing of the input, architecture and output of the neural
  network. See the text for details.}
  \label{fig:neural network}
\end{figure}

The neural network architecture depicted in \autoref{fig:neural network} is set up as follows. The input is a vector of 202
components: distance to the core $r$, secant of the zenith angle $\st$ and the
200 values of each trace $S_1,S_2,\dots,S_{200}$, where $S_i$ is the value of
the signal measured at the time bin $t_i$. At the start, this input is split into
two sets. One of them is the set of the time-independent variables, i.e.~$r$ and
$\st$. These variables are fed into a set of fully-connected layers that will compute the initial hidden state and cell state for the 
first layer of LSTMs. The fully-connected layers transform the inital vector of two values ($r$, $\st$) into a vector of dimension 32 and then to a vector of dimension 70. The outputs of the fully-connected layers together with the 200 time
values of each trace form the input to the LSTMs. Starting from a single sequence of 200 numbers (the signal $S_i$), the LSTMs produce 70, 32 and
32 sequences of 200 numbers, and the last one of these sequences is fed to a
final fully-connected layer. The activation function used for the fully connected layer is $f(x)=\max(0,x)$ and $\tanh$ for the LSTM layers. The model has a total of 87 212 trainable or free parameters.

The set of fully-connected layers computes the vector of initial parameters that encodes
information about the amount of atmosphere crossed by the particles, which in turn
changes the shape of the traces. Without this block, there is insufficient information for the neural network to distinguish between traces with high or low
fractions of muons and will be biased: for example, it will overestimate  the
muon component for nearly vertical showers and underestimate it for
more inclined showers. The block of LSTMs is responsible for computing the
traces and using the temporal information of the input.

\subsection{Data selection and training}
The neural network is trained with a library of simulated showers, with the same number of showers initiated by protons, helium, oxygen and iron nuclei. The energy range covered
by simulations spans from $10^{18.5}$\,eV to $10^{20.2}$ eV for zenith angles up
to $60^\circ$. Each shower is simulated with the CORSIKA software~\cite{Heck:1998vt} using the EPOS-LHC hadronic interaction model~\cite{Pierog:2013ria} and then reconstructed using the \Offline software of the Pierre Auger Collaboration~\cite{Argiro:2006cg}.
Among those simulated events, we select those that fulfill the quality condition that the station 
measuring the highest signal has to be surrounded by six operating neighbours, 
thus avoiding events that fall at the edges of the array. The method is applied to
traces whose signals do not show any sign of saturation caused by an overflow
of the read-out electronics or a loss of the linear behaviour of the PMT. Finally, only traces for
which the integral is more than 5\,VEM are included.

A total of
around 450\,000 events were available. The events were sampled randomly and assigned
to the training, validation or test data sets using a uniform distribution in
energy and $\st$ for the validation and test sets; the remaining events were assigned to
the training set. The training data set does not require special care regarding
the energy and zenith angle distributions since the dependence on the energy is
mild and the zenith angle is given as input. The whole event sample was split as
follows: around 390\,000 in the training set, 22\,000 in the validation set and
34\,000 in the test set.


Before training, both $r$ and $\st$ values are scaled to be between 0 and 1 and
all the traces are scaled individually to be between 0 and 1. The true (simulated) muon trace is scaled with the same factor used for the total trace. The output of the neural
network is also between 0 and 1 so as to use the same factor to rescale back the
predicted trace. The loss function that is minimized in the process of training is
the mean squared error, defined for the trace of a single WCD as
\begin{equation}
  L=\frac{1}{200}\sum_{i=1}^{200}\left(\smuptemp-\smuti\right)^2.
  \label{eq:loss}
\end{equation}
It corresponds to the average of the squares of the differences between the true muon
trace $\smuti$ and the predicted muon trace\footnote{In this work, we use a hat
  $\ \widehat{}\ $ for all the quantities that are either predicted by the neural
  network or computed from predictions of the neural network. For example, the integral of the muon signal is $S^\mu$, while the
  integral of the predicted muon signal is $\widehat{S^\mu}$.}
$\smuptemp$, for each time bin $i$. The neural network is trained
in batches and for each batch, the value of $L$ is computed for each trace and
averaged over all the samples in the batch.

\begin{figure}[t]
  \centering
  \includegraphics[width=.95\textwidth]{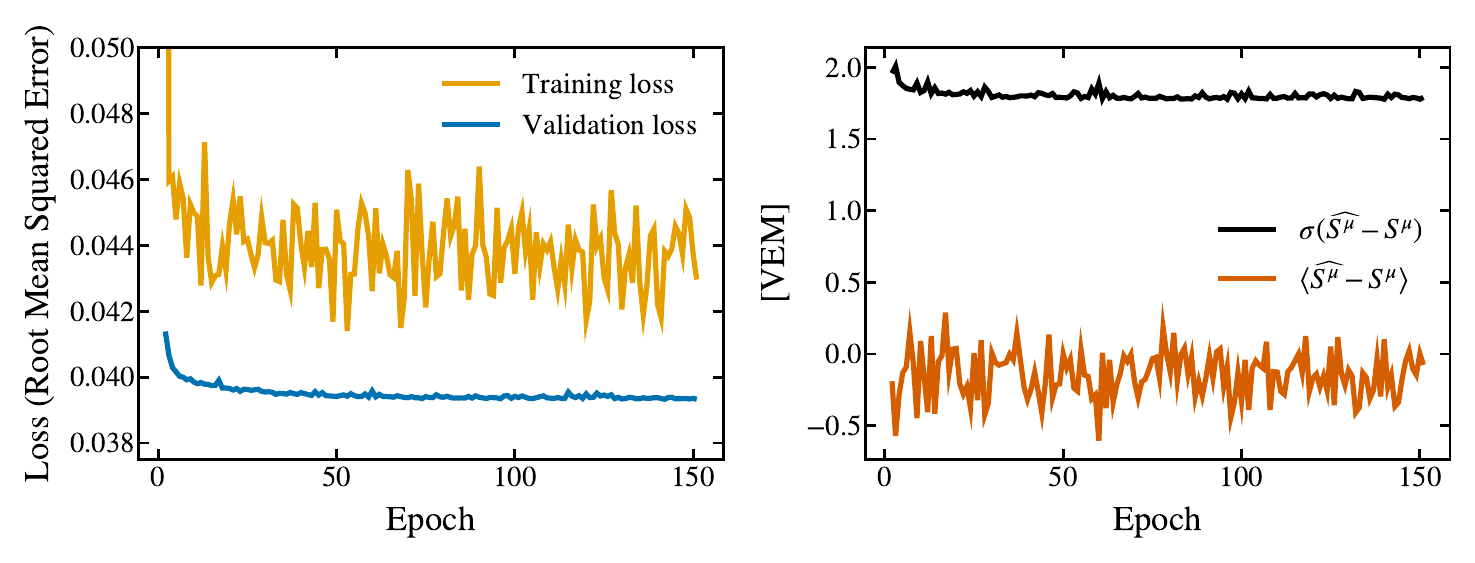}
  \caption{Left: Loss as a function of the epoch, see~\autoref{eq:loss}.
    Right: Mean value and standard deviation of the difference between the
    integral of the true muon signal and the predicted muon signal for the
    validation set.}
  \label{fig:loss}
\end{figure}

The training was done with the optimizer ADAM \cite{Kingma:2014vow} with a fixed learning
rate of $10^{-4}$ and the default values for the rest of the parameters; 
see ref.~\cite{Kingma:2014vow}. Using a batch size of 516 and 150 complete iterations over the whole training set or epochs on an Nvidia 2080 Ti GPU,
the training takes around 8 hours. The loss as a function of the epoch is shown
in \autoref{fig:loss} for both the training and validation sets. We can see that
the curve for the validation set decreases as the epoch increases. The curve for
validation is below the one for training because, as explained before, a uniform
distribution has been used for the validation data set and there are more events
for which the performance is worse (lower zenith angles) in the training set. To further assess the goodness of our approach, the difference between the integral of the
predicted and true muon traces is computed after each epoch. In the right panel
of \autoref{fig:loss}, we show the mean and the standard deviation of the
distribution of the difference. We use this measure because it has a
straightforward physical interpretation. The mean controls the bias: i.e., if the
mean is above zero, the neural network is consistently overestimating and vice
versa. The standard deviation tells us how well we are predicting the muon
signal, although this depends on many factors such as $\theta$,
as we will see in \autoref{sec:results}. All the pipeline was implemented in
Python 3.8 using Numpy \cite{2020NumPy-Array} and Pandas \cite{mckinney-proc-scipy-2010} for the data treatment and Pytorch
1.5.0 \cite{paszke2017automatic} for the construction and training of the neural
network.

Several tests were done to improve the performance of the method. Instead of
predicting the average muon trace, we aimed at predicting the muon signal for
each single PMT. These three independent results were averaged and compared to
the main method (i.e., feeding the network with the average of the signals of
the active PMTs). We found that even though the number of traces nearly
tripled, the performance did not improve, with an increase of 0.2~VEM in the
resolution or, in other words, around 1\% with respect to the total signal (see 
the bottom right panel of \autoref{fig:dif energy}). We also tried other changes 
in the neural network such as increasing the number of layers or using a decaying 
learning rate, with no significant improvements in performance.

\section{Results}
\label{sec:results}

\begin{figure}[t]
  \centering
\includegraphics[width=.95\textwidth]{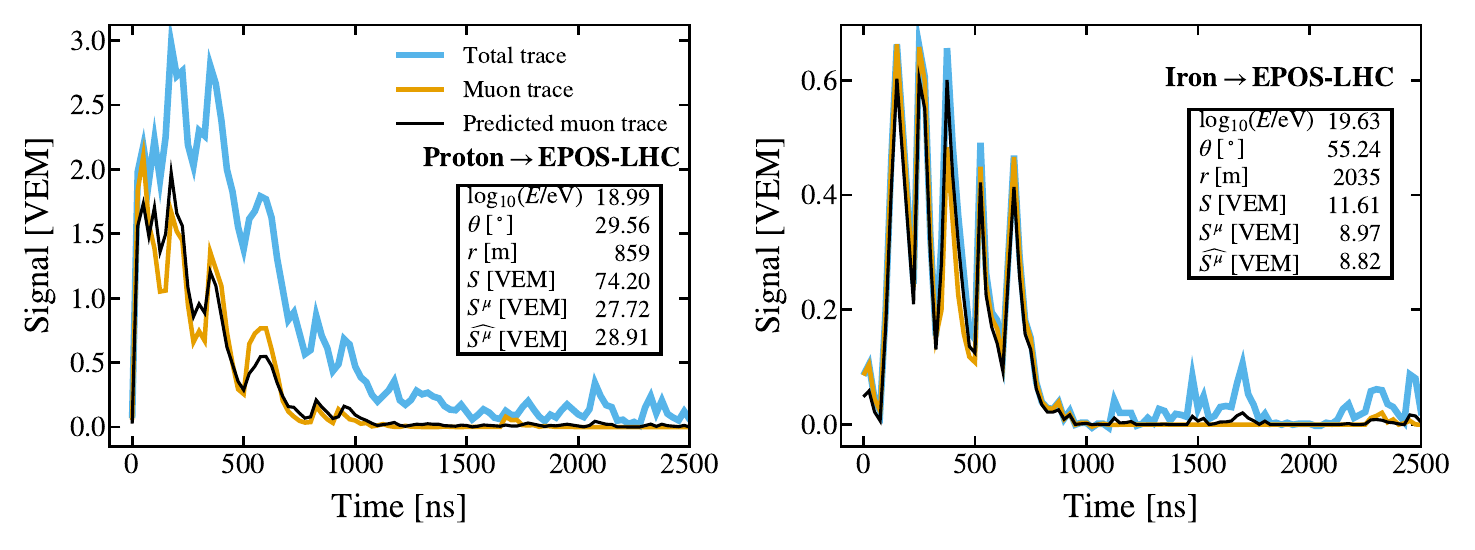}

  \caption{Examples of predicted muon traces for two simulated events with
    EPOS-LHC, for a electromagnetic-dominated signal (left) and muon-dominated signal (right). 
    The prediction (black line) agrees well with the shape of the simulated
    muon trace (orange line) for a majority of the time bins. The blue
    thicker line corresponds to the total trace, the one measured by
    a WCD.}
  \label{fig:traces}
\end{figure}

We show examples of the predictions of the neural network in \autoref{fig:traces}. As a general
comment, we observe that, qualitatively, the prediction follows the shape and
peaks of the total signal. The network has learnt to reproduce the main features
of the muon trace: its spiky shape and the fact that most muons arrive earlier.
A thorough discussion of the results and their dependence on several variables
are given in the next sections.

\subsection{Integrals of the trace}

\begin{figure}[t]
  \centering
  \includegraphics[width=.95\textwidth]{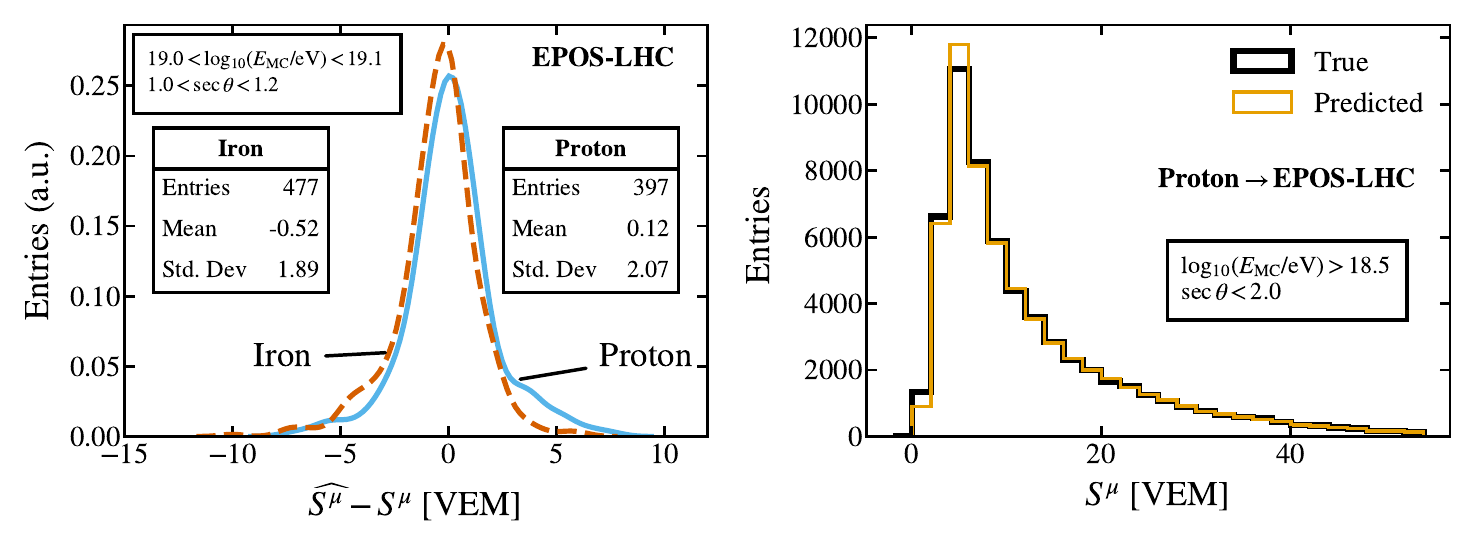}
  \caption{Left: Distribution of the difference between the integral of the
    predicted muon signal $\smup$ and the integral of the true muon signal
    $\smut$. Right: Distribution of $\smup$ and $\smut$ for all
    the stations in the test set.}
  \label{fig:differences-total-signal}
\end{figure}

One way to assess the performance of the method is to compute the integral of
the predicted muon trace, $\smup=\sum_{i=1}^{200}\smuptemp$
and compare it to the integral of the true muon trace
$\smut=\sum_{i=1}^{200}\smuti$. The integral of the muon trace is
an interesting physical observable, since it relates to the total number of
muons that reach the ground. In the left panel of
\autoref{fig:differences-total-signal}, we show a distribution with the
difference between $\smup$ and $\smut$ for a particular bin of energy and zenith
angle. The difference is compatible with zero and does not show a strong
dependence with the value of the true muon signal. In the right panel of
\autoref{fig:differences-total-signal}, we show the distribution of $\smup$
and $\smut$ for all the WCDs in the test set for showers initiated by a proton primary. 
The two distributions have similar shapes.

\begin{figure}[t]
\centering
\includegraphics[width=.95\textwidth]{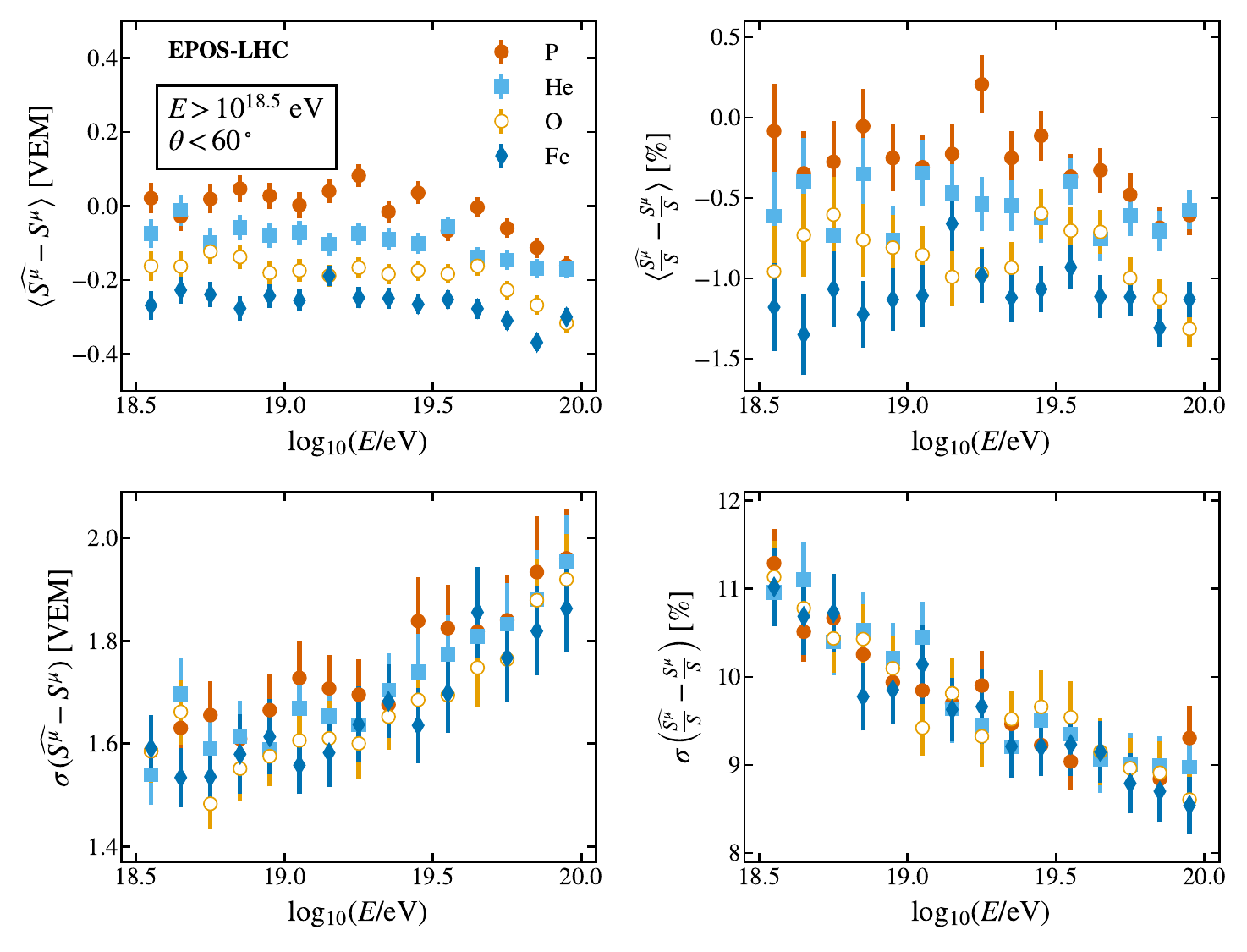}
\caption{Left: Mean value (top) and standard deviation (bottom) of the difference between the 
  predicted and values of the true muon signal as a function of energy. Right: 
  Relative bias (top) and resolution (bottom) for the determination of the muon fraction as a function of the energy.}
\label{fig:dif energy}
\end{figure}

\begin{figure}[t]
\includegraphics[width=.95\textwidth]{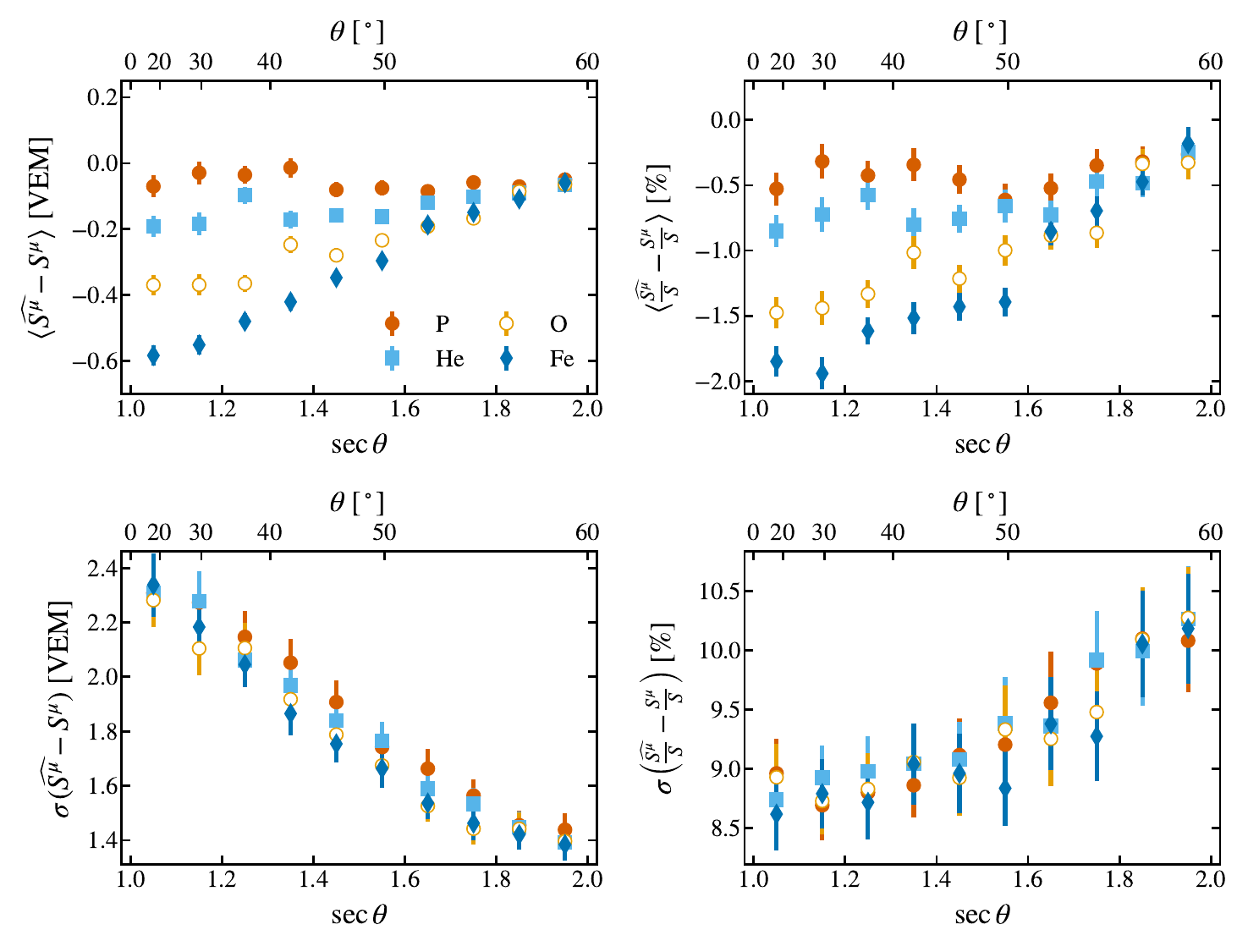}
\caption{Left: Mean value (top) and standard deviation (bottom) of the difference between the 
  predicted and values of the true muon signal as a function of $\st$. Right: 
  Relative bias (top) and resolution (bottom) for the determination of the muon fraction as a function of the zenith angle.}
\label{fig:dif zen}
\end{figure}

In \autoref{fig:dif energy} and \autoref{fig:dif zen}, the performances of the method 
are depicted as a function of the energy and $\st$, respectively. In the left
panels, the mean value and standard deviation of the difference between the 
predicted and true values are shown for the different primaries. The mean values
are close to zero and the standard deviation is less than 2\,VEMs in most bins. These
performances readily translate into the biases and resolutions for the muon fraction with respect to the total recorded signal shown in the right panels. The
biases are within 2\% for all the energies and angles explored here, regardless of the
primaries. The resolutions are of the order of 11\% at $10^{18.5}~$eV, improving
with higher energies. The dependence on $\st$ reflects the attenuation of the overall
signal at large zenith angles.

\begin{figure}[t]
\centering
\includegraphics[width=.95\textwidth]{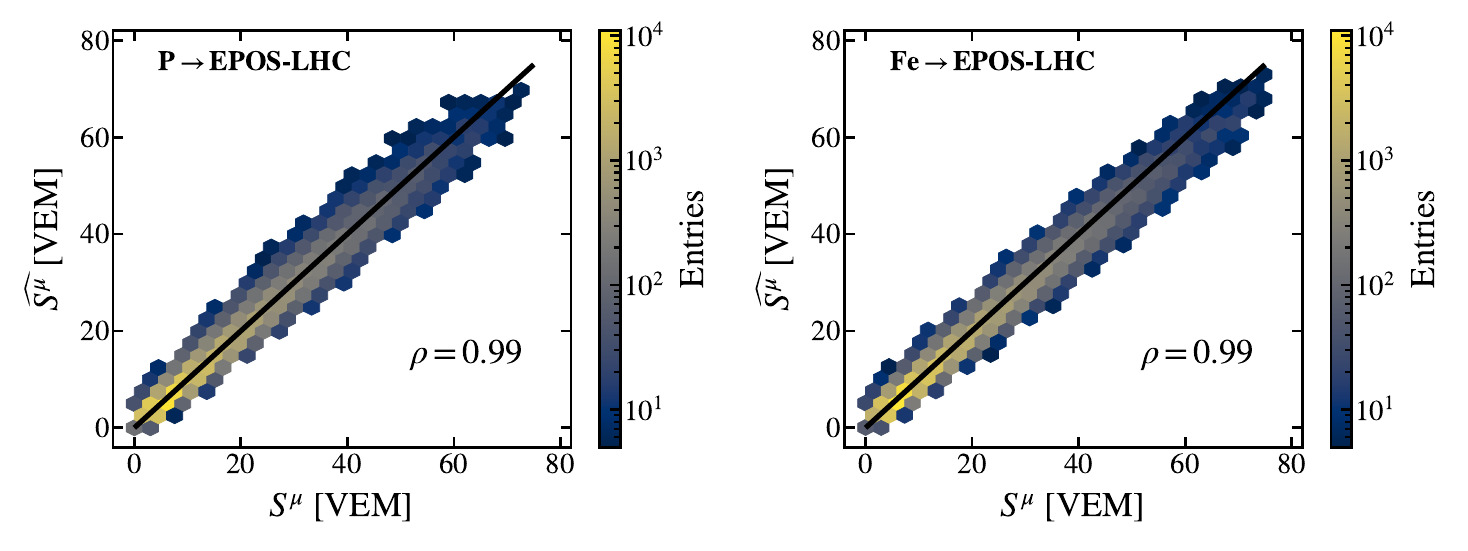}
\caption{Integral of the predicted muon trace as a function of the integral of
  the true muon trace. The black line corresponds to a perfect prediction and
  $\rho$ is the Pearson correlation coefficient.}
\label{fig:correlation paper}
\end{figure}

\begin{figure}[t]
  \centering
  \includegraphics[width=.95\textwidth]{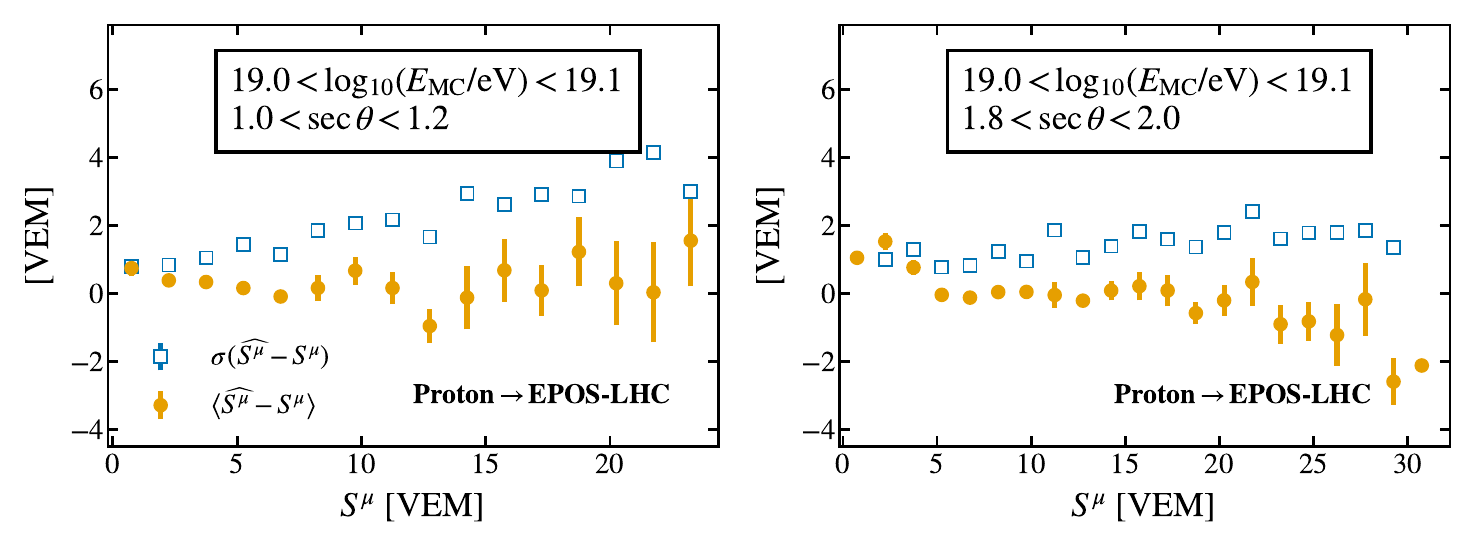}
  \caption{Mean and standard deviation of the difference between the integral of
    the predicted muon signal $\smup$ and the integral of the true muon signal
    $\smut$ for all the stations from events with the energies and zenith angles
    specified in the boxes.}
  \label{fig:mean-std-total-signal}
\end{figure}

The performance of the method can be also evaluated as a function of the simulated muon signal ($\smut$). In \autoref{fig:correlation paper}, we show the density
plot of the predictions and the true values. The predictions are highly correlated 
with a Pearson correlation coefficient that is 0.99 for all the primaries. The mean 
and standard deviation of the distribution of $\smup-\smut$ are shown in \autoref{fig:mean-std-total-signal} as a function of $\smut$. The mean is close 
to zero and rarely exceeds 2\,VEM. These results are observed to be zenith-angle dependent. The 
standard deviation increases as $\smut$ increases for more vertical events (left panel), 
while it remains constant at large angles (right panel) as a result of the
preponderance of the muon signal.

For $\smut > 5$ VEM, the resolution in the determination of the muon signal for each individual WCD goes from about 20\% for the case of vertical events to 10\% for events with $\st \simeq 2.0$

\subsection{Risetime of the muon signal}
The risetime of the recorded signals is another observable worth analysing,
since it has been used successfully by the Pierre Auger Collaboration to extract
valuable information about extensive air showers \cite{Aab:2016enk,Aab:2017cgk}. The
risetime $t_{1/2}$ is defined as the difference between two times, $t_{10}$ and
$t_{50}$. $t_{10}$ is the time corresponding to the first time bin where the integral of the signal reaches 10 \% of its total value, while
$t_{50}$ is obtained when the integral reaches 50 \% of the total. The risetime gives information about the shape of the
trace: traces where the signal is concentrated in a few bins will have a shorter
risetime, while traces where the signal is spread over time will have larger
risetimes. In particular, the muon component has a smaller risetime than the
electromagnetic component, since muons arrive earlier and in a shorter window of
time \cite{1962PhRv..128.2384L}.

\begin{figure}[t]
  \centering
  \includegraphics[width=.95\textwidth]{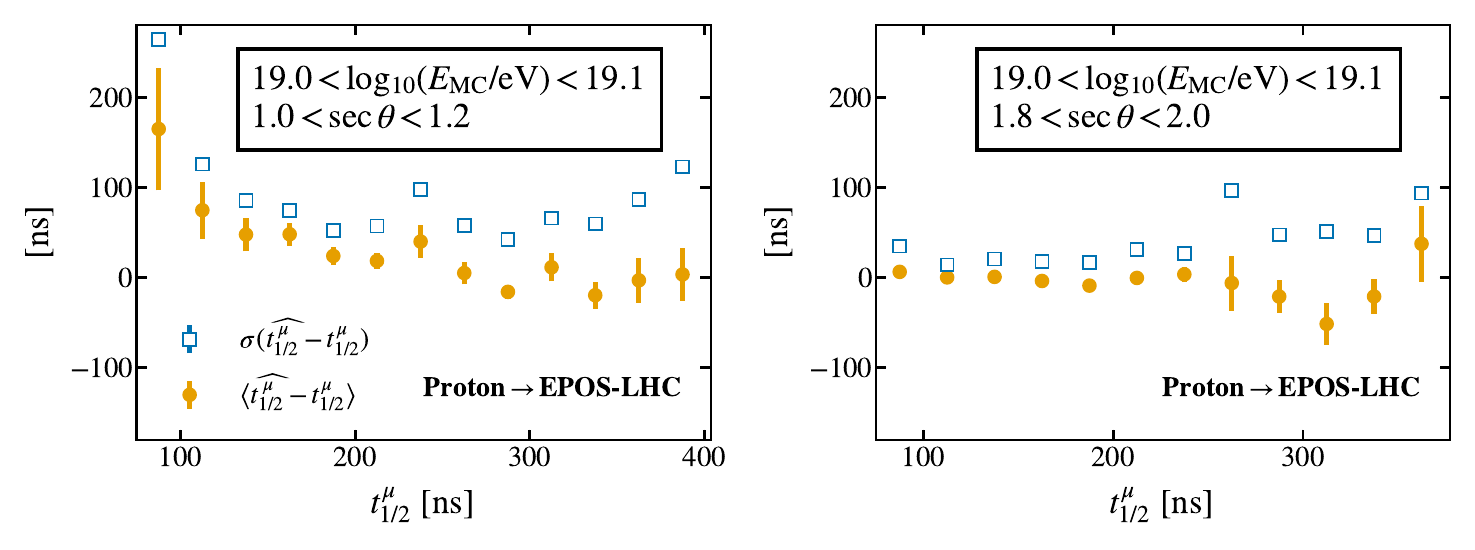}
  \caption{Mean and standard deviation of the difference between the risetime of
    the predicted muon signal $\rtmp$ and the risetime of the true muon signal
    $\rtm$ for all the stations from events with the energies and zenith angles
    specified in the boxes.}
  \label{fig:mean-std-risetime}
\end{figure}

In \autoref{fig:mean-std-risetime}, we compare the risetime of the predicted muon
trace $\rtmp$ with the risetime of the simulated muon trace $\rtm$. The standard
deviation is less than 100\,ns (4 time bins) for most values of the true risetime. This is a
small time compared to the risetime of the total signal, which has a mean of 500 ns and 150 ns for the left and rights bins of \autoref{fig:mean-std-risetime}, respectively. Values with $\rtm<100$ ns correspond to around 1.5\% of the samples in the left panel of \autoref{fig:mean-std-risetime}, since for vertical showers it is very rare to have the muon signal concentrated in very few bins, while for inclined showers it is rare that the muon signal is spread over a large time. The mean value approaches zero and the performance
improves as the zenith angle increases. This means that the network successfully
predicts not only the integral of the muon trace but also the shape of the muon
trace.

\subsection{Hadronic interaction model}
We have chosen to train our neural network on simulations with the EPOS-LHC 
generator of hadronic interactions and 
presented some tests in the previous sections using the same event generator. We now test our method using 
simulations done with QGSJetII-04~\cite{Ostapchenko:2010vb} and Sibyll 2.3~\cite{Ahn:2009wx} event generators. 
In other words, we test the predictions for simulations that are not only unknown 
to the neural network but that also have been generated using a different model of
hadronic interactions. 

\begin{figure}[t]
  \centering
  \includegraphics[width=.95\textwidth]{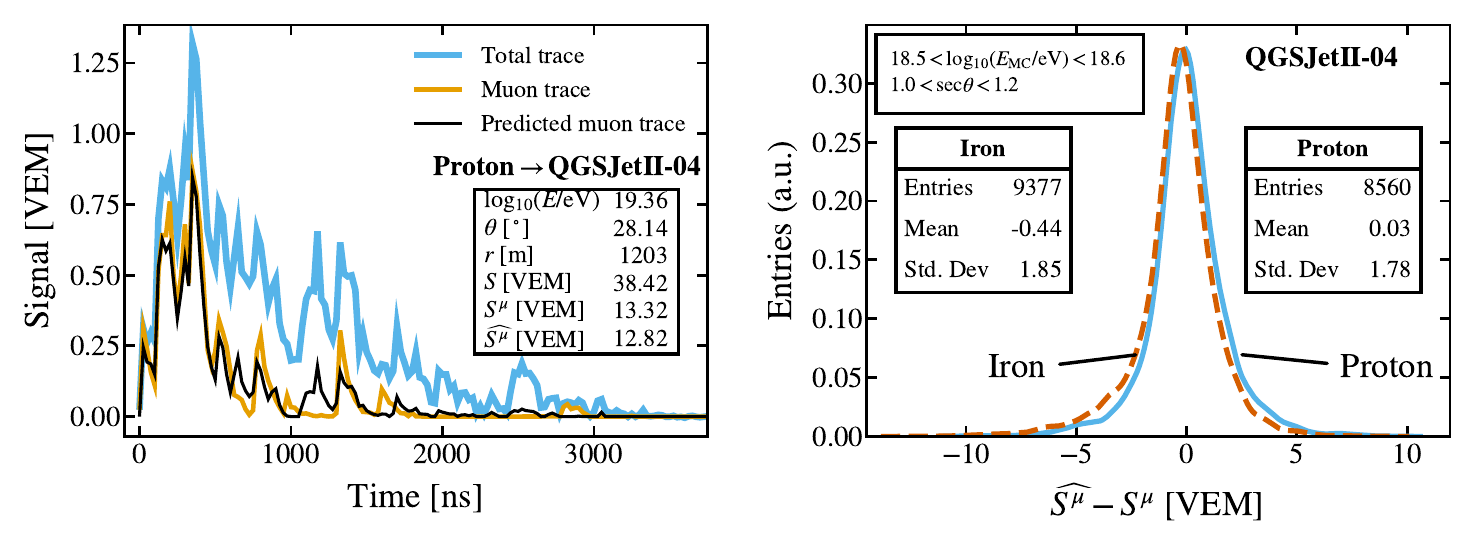}
  \caption{Left: Example of a predicted trace for a simulation of a proton
    generated air shower done with QGSJetII-04. Right: Distribution of
    $\smup-\smut$ for all the stations in the bin specified for simulations
    using proton and iron nuclei.}
  \label{fig:qgs-trace}
\end{figure}

\paragraph{QGSJetII-04.}
In \autoref{fig:qgs-trace}, an example of a trace obtained with
simulations produced using QGSJetII-04 is shown. The prediction follows the shape of the
peaks, predicting quite accurately the muon signal. The difference between true
and predicted muon signals does not show a strong deviance from zero.

\begin{figure}[t]
  \centering
  \includegraphics[width=.95\textwidth]{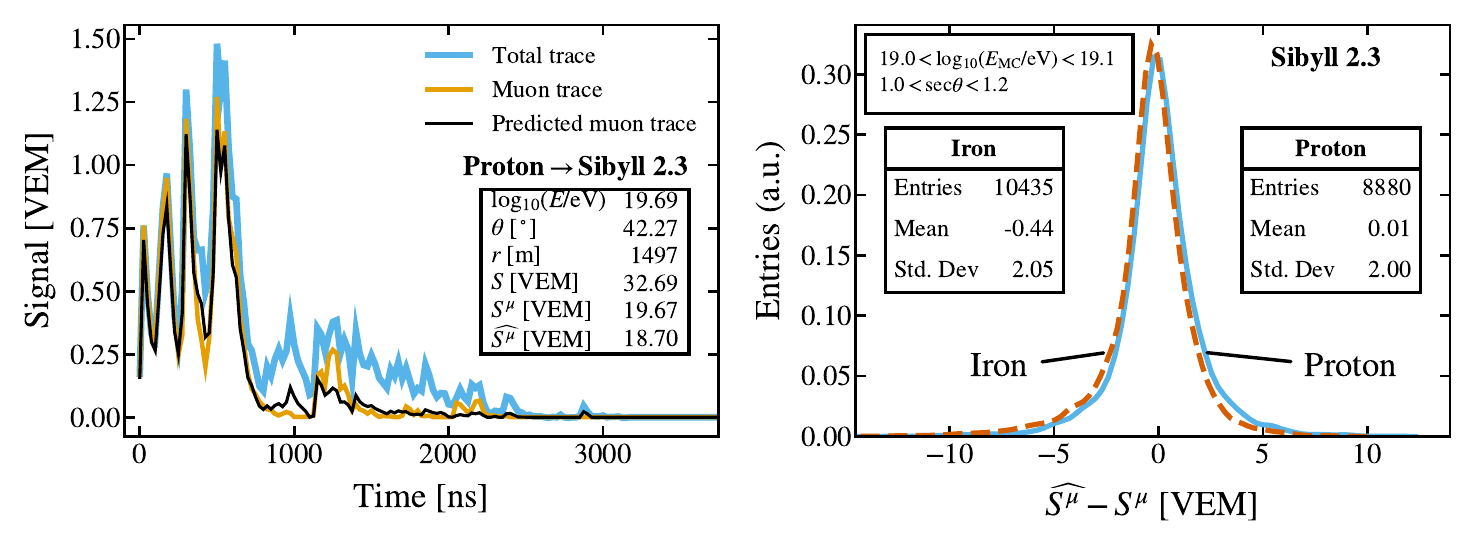}
  \caption{Left: Example of a predicted trace for a simulation done with
    Sibyll 2.3 with a proton as primary cosmic ray. Right: Distribution of
    $\smup-\smut$ for all the stations in the bin specified for simulations using
    proton and iron nuclei.}
  \label{fig:sib trace}
\end{figure}

\paragraph{Sibyll 2.3.}
In \autoref{fig:sib trace}, we show the result of predicting the muon traces for
simulations performed with Sibyll 2.3. By comparing the values of the bias and of the
resolution, we can see that the performance is similar to the case where the
predictions were based on simulations done with QGSJetII-04.

With these results we show that the neural network predictions rely essentially on 
the response of the detectors to muons and are thus independent of the
hadronic interaction model used to simulate extensive air showers. For completeness, we have
also carried out the opposite exercise: train the neural network with
QGSJetII-04 and Sibyll 2.3 and predict for the other two models, respectively. The
outcome is in a good agreement with the case where the
neural network learns from events simulated with EPOS-LHC.

\begin{figure}[th]
  \centering
  \includegraphics[width=.95\textwidth]{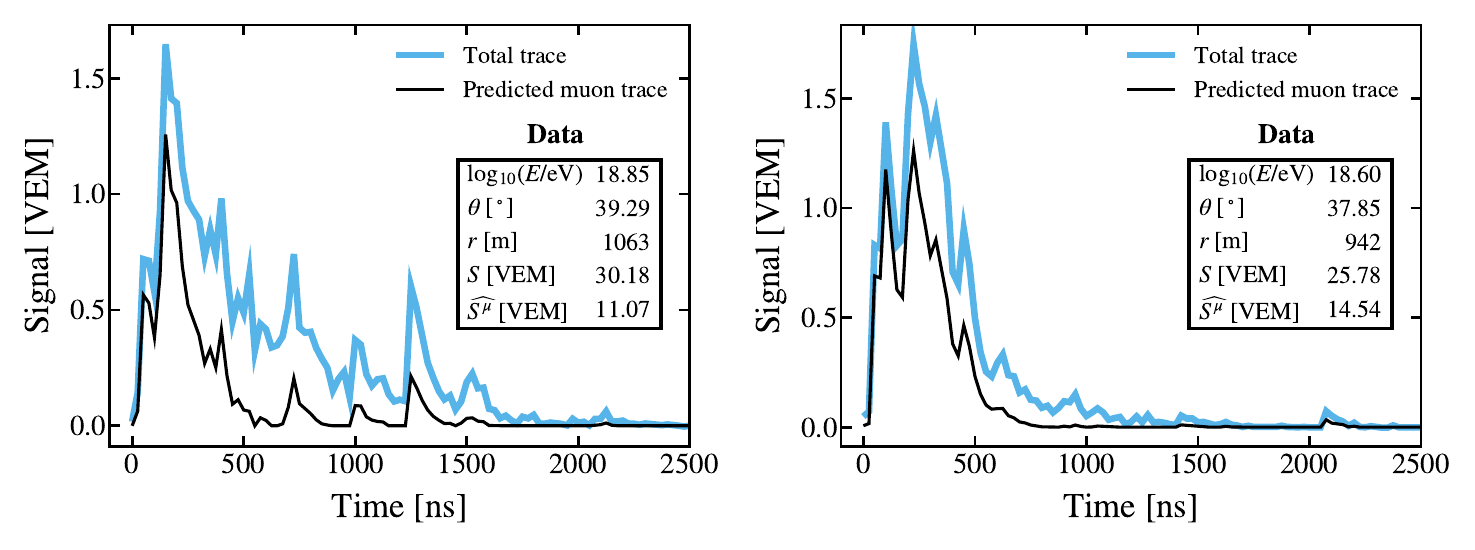}
  \caption{Examples of the predicted muon traces for two WCDs that belong to
    two different events recorded by the SD.}
  \label{fig:traces-data}
\end{figure}

\subsection{Comparison to data}
In this section we assess how the neural network performs
when applied to experimental data. The sample of events selected have the same cuts as the simulations used for training. In total, there are 177\,000 events registered from 2004 to 2017. In \autoref{fig:traces-data}, we show examples
of the muon trace predicted for two typical signals recorded with two independent WCDs.  We
observe that the output of the neural network produces features similar to those
shown by the simulated traces of \autoref{fig:traces}, namely: the predicted muon
signal is larger at earlier times since muons arrive earlier; in addition, the
predicted muon trace exhibits a spiky structure.

\begin{figure}[th]
  \centering
  \includegraphics[width=.47\textwidth]{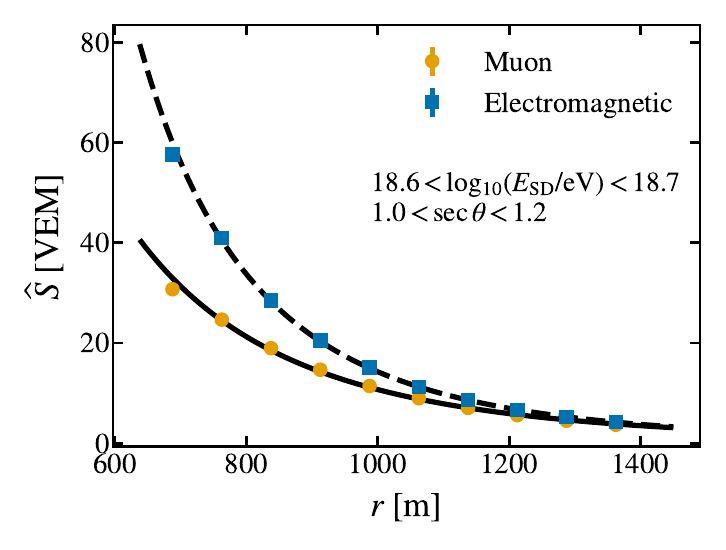}
  \includegraphics[width=.47\textwidth]{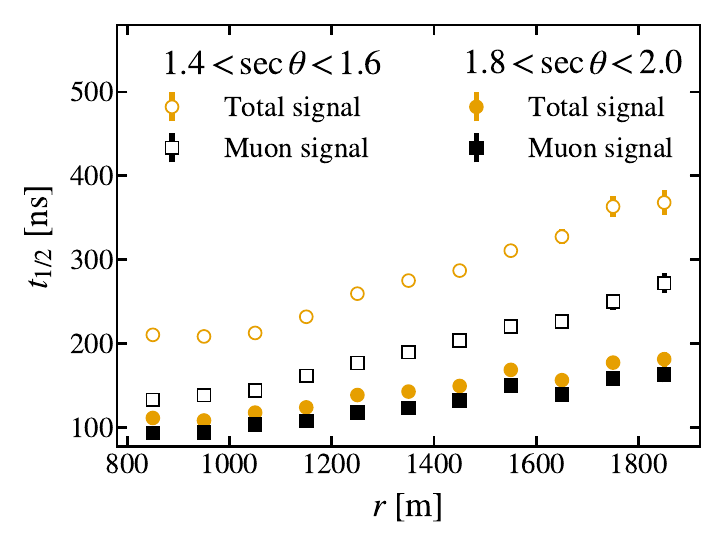}
  \caption{Left: 
    Lateral distribution of the predicted muon and electromagnetic
    components for data. The solid and dashed lines are the outcome of two
    independent fits (see text for details). 
    Right: Measured risetime of the total signal and  risetime of the predicted muon signal as a function of the distance for two different zenith angle ranges.
    }
  \label{fig:muon_deficit_and_rt_data}
\end{figure}

A straightforward way to verify the robustness of the muon signal measurement is
to study the behaviour of the total muon signal with the distance to the air-shower axis, i.e.~the muon lateral distribution function (LDF). The muon LDF is illustrated in~\autoref{fig:muon_deficit_and_rt_data}, together with the electromagnetic one. For each WCD, we can obtain the electromagnetic component
simply by subtracting the predicted muon signal from the total recorded signal. This endows us with the additional possibility to study the behaviour of the
LDF of electromagnetic signals.

In addition, we compare our data to the parameterizations of the LDF of
muon and electromagnetic particles that best fit the Akeno measurements. The
authors of ref.~\cite{Hayashida:1995tu} studied the properties of muons above 1\,GeV at
different distances to the shower core for events with $\sec \theta < 1.2$. The
behaviour of the Akeno data is well reproduced by the function proposed by
Greisen in ref.~\cite{Greisen}, Eqs.~(4) and (6) in ref.~\cite{Hayashida:1995tu}. We fit our
data with those expressions and the parameters that best reproduce Akeno data,
just leaving the overall normalization free. The outcome of the fit is shown as a solid line in~\autoref{fig:muon_deficit_and_rt_data} for the energy range
$10^{18.6}$\,eV to $10^{18.7}$\,eV. The level of agreement between our measurements and
the muon lateral distribution function extracted from Akeno data is remarkable.

In ref.~\cite{Nagano:1991jz}, the
Akeno data are used to extract a parameterization of the density of the
electromagnetic signal as a function of the distance to the shower core
(Eq.~(2.1) in ref.~\cite{Nagano:1991jz}). We use that parameterization to fit our
predicted electromagnetic signal, leaving the overall normalization free and the
parameter referred to as $\eta$ in the original publication. $\eta$  is left as a free parameter since it is a function of the measured electromagnetic signal (see Eq.~(2.2) in ref.~\cite{Nagano:1991jz}). As for the case of muons, the agreement with the Akeno
parameterization of the electromagnetic signals is very good (dashed line in \autoref{fig:muon_deficit_and_rt_data}).

The data sample was also used to carry out a cross-check involving the arrival times of the particles: we compared the behaviour with the distance to the core and with the zenith angle of the risetimes measured for the total signal and the predicted muon signal, see the right panel of~\autoref{fig:muon_deficit_and_rt_data}. As expected, the muon risetime is consistently smaller since muons arrive earlier. As the zenith angle increases, the fraction of the muon signal grows, thus the risetime of the total signal and the muon risetime become more similar. This behaviour is correctly reproduced by the predictions of the RNN. 



\section{Conclusions}
\label{sec:conclusions}
We have shown that over a broad range of energies and zenith angles,
a Recurrent Neural Network is a suitable tool to predict
the muon signals that are part
of the time traces measured with each individual WCD of the SD of the Pierre Auger
Observatory. The neural network is trained on simulations, but the predictions
are independent of the model used to simulate hadronic interactions. The muon signal for each WCD can be measured with a resolution that decreases from 20\% to 10\% as the zenith angle increases. Likewise, the muon fraction with respect to the total signal is estimated with biases that are within a 2\% and a resolution that is lower than 11\%.

When applied to data, the behaviour of the extracted muon and electromagnetic
signals agrees well with relevant measurements found in the literature while other observables, such as the risetime, follow the behaviour dictated by basic physics principles.

The combination of this algorithm with the future data collected with the upgraded
Observatory~\cite{Aab:2016vlz} will represent a major step forward since arguably, we will
achieve an unprecedented resolution in our capability to make mass estimates on
an event-by-event basis.

\input{acknowledgments}
\begin{sloppypar}
The authors gratefully acknowledge the support of NVIDIA Corporation with the donation of GPU hardware used for this research.
\end{sloppypar}

\bibliographystyle{JHEP.bst}
\bibliography{bibliography.bib}

\begin{center}
\rule{0.1\columnwidth}{0.5pt}\,\raisebox{-0.5pt}{\rule{0.05\columnwidth}{1.5pt}}~\raisebox{-0.375ex}{\scriptsize$\bullet$}~\raisebox{-0.5pt}{\rule{0.05\columnwidth}{1.5pt}}\,\rule{0.1\columnwidth}{0.5pt}
\end{center}

\section*{The Pierre Auger Collaboration}
\input{latex_authorlist_authors}
{\footnotesize
\input{latex_authorlist_institutions}
}
 
\end{document}

%% file: acknowledgments.tex

\section*{Acknowledgments}

\begin{sloppypar}
The successful installation, commissioning, and operation of the Pierre
Auger Observatory would not have been possible without the strong
commitment and effort from the technical and administrative staff in
Malarg\"ue. We are very grateful to the following agencies and
organizations for financial support:
\end{sloppypar}

\begin{sloppypar}
Argentina -- Comisi\'on Nacional de Energ\'\i{}a At\'omica; Agencia Nacional de
Promoci\'on Cient\'\i{}fica y Tecnol\'ogica (ANPCyT); Consejo Nacional de
Investigaciones Cient\'\i{}ficas y T\'ecnicas (CONICET); Gobierno de la
Provincia de Mendoza; Municipalidad de Malarg\"ue; NDM Holdings and Valle
Las Le\~nas; in gratitude for their continuing cooperation over land
access; Australia -- the Australian Research Council; Brazil -- Conselho
Nacional de Desenvolvimento Cient\'\i{}fico e Tecnol\'ogico (CNPq);
Financiadora de Estudos e Projetos (FINEP); Funda\c{c}\~ao de Amparo \`a
Pesquisa do Estado de Rio de Janeiro (FAPERJ); S\~ao Paulo Research
Foundation (FAPESP) Grants No.~2019/10151-2, No.~2010/07359-6 and
No.~1999/05404-3; Minist\'erio da Ci\^encia, Tecnologia, Inova\c{c}\~oes e
Comunica\c{c}\~oes (MCTIC); Czech Republic -- Grant No.~MSMT CR LTT18004,
LM2015038, LM2018102, CZ.02.1.01/0.0/0.0/16{\textunderscore}013/0001402,
CZ.02.1.01/0.0/0.0/18{\textunderscore}046/0016010 and
CZ.02.1.01/0.0/0.0/17{\textunderscore}049/0008422; France -- Centre de Calcul
IN2P3/CNRS; Centre National de la Recherche Scientifique (CNRS); Conseil
R\'egional Ile-de-France; D\'epartement Physique Nucl\'eaire et Corpusculaire
(PNC-IN2P3/CNRS); D\'epartement Sciences de l'Univers (SDU-INSU/CNRS);
Institut Lagrange de Paris (ILP) Grant No.~LABEX ANR-10-LABX-63 within
the Investissements d'Avenir Programme Grant No.~ANR-11-IDEX-0004-02;
Germany -- Bundesministerium f\"ur Bildung und Forschung (BMBF); Deutsche
Forschungsgemeinschaft (DFG); Finanzministerium Baden-W\"urttemberg;
Helmholtz Alliance for Astroparticle Physics (HAP);
Helmholtz-Gemeinschaft Deutscher Forschungszentren (HGF); Ministerium
f\"ur Innovation, Wissenschaft und Forschung des Landes
Nordrhein-Westfalen; Ministerium f\"ur Wissenschaft, Forschung und Kunst
des Landes Baden-W\"urttemberg; Italy -- Istituto Nazionale di Fisica
Nucleare (INFN); Istituto Nazionale di Astrofisica (INAF); Ministero
dell'Istruzione, dell'Universit\'a e della Ricerca (MIUR); CETEMPS Center
of Excellence; Ministero degli Affari Esteri (MAE); M\'exico -- Consejo
Nacional de Ciencia y Tecnolog\'\i{}a (CONACYT) No.~167733; Universidad
Nacional Aut\'onoma de M\'exico (UNAM); PAPIIT DGAPA-UNAM; The Netherlands
-- Ministry of Education, Culture and Science; Netherlands Organisation
for Scientific Research (NWO); Dutch national e-infrastructure with the
support of SURF Cooperative; Poland -Ministry of Science and Higher
Education, grant No.~DIR/WK/2018/11; National Science Centre, Grants
No.~2013/08/M/ST9/00322, No.~2016/23/B/ST9/01635 and No.~HARMONIA
5--2013/10/M/ST9/00062, UMO-2016/22/M/ST9/00198; Portugal -- Portuguese
national funds and FEDER funds within Programa Operacional Factores de
Competitividade through Funda\c{c}\~ao para a Ci\^encia e a Tecnologia
(COMPETE); Romania -- Romanian Ministry of Education and Research, the
Program Nucleu within MCI (PN19150201/16N/2019 and PN19060102) and
project PN-III-P1-1.2-PCCDI-2017-0839/19PCCDI/2018 within PNCDI III;
Slovenia -- Slovenian Research Agency, grants P1-0031, P1-0385, I0-0033,
N1-0111; Spain -- Ministerio de Econom\'\i{}a, Industria y Competitividad
(FPA2017-85114-P and PID2019-104676GB-C32, Xunta de Galicia (ED431C
2017/07), Junta de Andaluc\'\i{}a (SOMM17/6104/UGR, P18-FR-4314) Feder Funds,
RENATA Red Nacional Tem\'atica de Astropart\'\i{}culas (FPA2015-68783-REDT) and
Mar\'\i{}a de Maeztu Unit of Excellence (MDM-2016-0692); USA -- Department of
Energy, Contracts No.~DE-AC02-07CH11359, No.~DE-FR02-04ER41300,
No.~DE-FG02-99ER41107 and No.~DE-SC0011689; National Science Foundation,
Grant No.~0450696; The Grainger Foundation; Marie Curie-IRSES/EPLANET;
European Particle Physics Latin American Network; and UNESCO.
\end{sloppypar}

%% file: latex_authorlist_authors.tex

A.~Aab$^{80}$,
P.~Abreu$^{72}$,
M.~Aglietta$^{52,50}$,
J.M.~Albury$^{12}$,
I.~Allekotte$^{1}$,
A.~Almela$^{8,11}$,
J.~Alvarez-Mu\~niz$^{79}$,
R.~Alves Batista$^{80}$,
G.A.~Anastasi$^{61,50}$,
L.~Anchordoqui$^{87}$
% ...

%% file: latex_authorlist_institutions.tex

\begin{description}[labelsep=0.2em,align=right,labelwidth=0.7em,labelindent=0em,leftmargin=2em,noitemsep]
\item[$^{1}$] Centro At\'omico Bariloche and Instituto Balseiro (CNEA-UNCuyo-CONICET), San Carlos de Bariloche, Argentina
\item[$^{2}$] Centro de Investigaciones en L\'aseres y Aplicaciones, CITEDEF and CONICET, Villa Martelli, Argentina
\item[$^{3}$] Departamento de F\'\i{}sica and Departamento de Ciencias de la Atm\'osfera y los Oc\'eanos, FCEyN, Universidad de Buenos Aires and CONICET, Buenos Aires, Argentina
\item[$^{4}$] IFLP, Universidad Nacional de La Plata and CONICET, La Plata, Argentina
\item[$^{5}$] Instituto de Astronom\'\i{}a y F\'\i{}sica del Espacio (IAFE, CONICET-UBA), Buenos Aires, Argentina
\item[$^{6}$] Instituto de F\'\i{}sica de Rosario (IFIR) -- CONICET/U.N.R.\ and Facultad de Ciencias Bioqu\'\i{}micas y Farmac\'euticas U.N.R., Rosario, Argentina
\item[$^{7}$] Instituto de Tecnolog\'\i{}as en Detecci\'on y Astropart\'\i{}culas (CNEA, CONICET, UNSAM), and Universidad Tecnol\'ogica Nacional -- Facultad Regional Mendoza (CONICET/CNEA), Mendoza, Argentina
\item[$^{8}$] Instituto de Tecnolog\'\i{}as en Detecci\'on y Astropart\'\i{}culas (CNEA, CONICET, UNSAM), Buenos Aires, Argentina
\item[$^{9}$] Observatorio Pierre Auger, Malarg\"ue, Argentina
\item[$^{10}$] Observatorio Pierre Auger and Comisi\'on Nacional de Energ\'\i{}a At\'omica, Malarg\"ue, Argentina
\item[$^{11}$] Universidad Tecnol\'ogica Nacional -- Facultad Regional Buenos Aires, Buenos Aires, Argentina
\item[$^{12}$] University of Adelaide, Adelaide, S.A., Australia
\item[$^{13}$] Universit\'e Libre de Bruxelles (ULB), Brussels, Belgium
\item[$^{14}$] Vrije Universiteit Brussels, Brussels, Belgium
\item[$^{15}$] Centro Brasileiro de Pesquisas Fisicas, Rio de Janeiro, RJ, Brazil
\item[$^{16}$] Centro Federal de Educa\c{c}\~ao Tecnol\'ogica Celso Suckow da Fonseca, Nova Friburgo, Brazil
\item[$^{17}$] Universidade de S\~ao Paulo, Escola de Engenharia de Lorena, Lorena, SP, Brazil
\item[$^{18}$] Universidade de S\~ao Paulo, Instituto de F\'\i{}sica de S\~ao Carlos, S\~ao Carlos, SP, Brazil
\item[$^{19}$] Universidade de S\~ao Paulo, Instituto de F\'\i{}sica, S\~ao Paulo, SP, Brazil
\item[$^{20}$] Universidade Estadual de Campinas, IFGW, Campinas, SP, Brazil
\item[$^{21}$] Universidade Estadual de Feira de Santana, Feira de Santana, Brazil
\item[$^{22}$] Universidade Federal do ABC, Santo Andr\'e, SP, Brazil
\item[$^{23}$] Universidade Federal do Paran\'a, Setor Palotina, Palotina, Brazil
\item[$^{24}$] Universidade Federal do Rio de Janeiro, Instituto de F\'\i{}sica, Rio de Janeiro, RJ, Brazil
\item[$^{25}$] Universidade Federal do Rio de Janeiro (UFRJ), Observat\'orio do Valongo, Rio de Janeiro, RJ, Brazil
\item[$^{26}$] Universidade Federal Fluminense, EEIMVR, Volta Redonda, RJ, Brazil
\item[$^{27}$] Universidad de Medell\'\i{}n, Medell\'\i{}n, Colombia
\item[$^{28}$] Universidad Industrial de Santander, Bucaramanga, Colombia
\item[$^{29}$] Charles University, Faculty of Mathematics and Physics, Institute of Particle and Nuclear Physics, Prague, Czech Republic
\item[$^{30}$] Institute of Physics of the Czech Academy of Sciences, Prague, Czech Republic
\item[$^{31}$] Palacky University, RCPTM, Olomouc, Czech Republic
\item[$^{32}$] CNRS/IN2P3, IJCLab, Universit\'e Paris-Saclay, Orsay, France
\item[$^{33}$] Laboratoire de Physique Nucl\'eaire et de Hautes Energies (LPNHE), Sorbonne Universit\'e, Universit\'e de Paris, CNRS-IN2P3, Paris, France
\item[$^{34}$] Univ.\ Grenoble Alpes, CNRS, Grenoble Institute of Engineering Univ.\ Grenoble Alpes, LPSC-IN2P3, 38000 Grenoble, France
\item[$^{35}$] Universit\'e Paris-Saclay, CNRS/IN2P3, IJCLab, Orsay, France
\item[$^{36}$] Bergische Universit\"at Wuppertal, Department of Physics, Wuppertal, Germany
\item[$^{37}$] Karlsruhe Institute of Technology (KIT), Institute for Experimental Particle Physics, Karlsruhe, Germany
\item[$^{38}$] Karlsruhe Institute of Technology (KIT), Institut f\"ur Prozessdatenverarbeitung und Elektronik, Karlsruhe, Germany
\item[$^{39}$] Karlsruhe Institute of Technology (KIT), Institute for Astroparticle Physics, Karlsruhe, Germany
\item[$^{40}$] RWTH Aachen University, III.\ Physikalisches Institut A, Aachen, Germany
\item[$^{41}$] Universit\"at Hamburg, II.\ Institut f\"ur Theoretische Physik, Hamburg, Germany
\item[$^{42}$] Universit\"at Siegen, Department Physik -- Experimentelle Teilchenphysik, Siegen, Germany
\item[$^{43}$] Gran Sasso Science Institute, L'Aquila, Italy
\item[$^{44}$] INFN Laboratori Nazionali del Gran Sasso, Assergi (L'Aquila), Italy
\item[$^{45}$] INFN, Sezione di Catania, Catania, Italy
\item[$^{46}$] INFN, Sezione di Lecce, Lecce, Italy
\item[$^{47}$] INFN, Sezione di Milano, Milano, Italy
\item[$^{48}$] INFN, Sezione di Napoli, Napoli, Italy
\item[$^{49}$] INFN, Sezione di Roma ``Tor Vergata'', Roma, Italy
\item[$^{50}$] INFN, Sezione di Torino, Torino, Italy
\item[$^{51}$] Istituto di Astrofisica Spaziale e Fisica Cosmica di Palermo (INAF), Palermo, Italy
\item[$^{52}$] Osservatorio Astrofisico di Torino (INAF), Torino, Italy
\item[$^{53}$] Politecnico di Milano, Dipartimento di Scienze e Tecnologie Aerospaziali , Milano, Italy
\item[$^{54}$] Universit\`a del Salento, Dipartimento di Matematica e Fisica ``E.\ De Giorgi'', Lecce, Italy
\item[$^{55}$] Universit\`a dell'Aquila, Dipartimento di Scienze Fisiche e Chimiche, L'Aquila, Italy
\item[$^{56}$] Universit\`a di Catania, Dipartimento di Fisica e Astronomia, Catania, Italy
\item[$^{57}$] Universit\`a di Milano, Dipartimento di Fisica, Milano, Italy
\item[$^{58}$] Universit\`a di Napoli ``Federico II'', Dipartimento di Fisica ``Ettore Pancini'', Napoli, Italy
\item[$^{59}$] Universit\`a di Palermo, Dipartimento di Fisica e Chimica ''E.\ Segr\`e'', Palermo, Italy
\item[$^{60}$] Universit\`a di Roma ``Tor Vergata'', Dipartimento di Fisica, Roma, Italy
\item[$^{61}$] Universit\`a Torino, Dipartimento di Fisica, Torino, Italy
\item[$^{62}$] Benem\'erita Universidad Aut\'onoma de Puebla, Puebla, M\'exico
\item[$^{63}$] Centro de Investigaci\'on y de Estudios Avanzados del IPN (CINVESTAV), M\'exico, D.F., M\'exico
\item[$^{64}$] Unidad Profesional Interdisciplinaria en Ingenier\'\i{}a y Tecnolog\'\i{}as Avanzadas del Instituto Polit\'ecnico Nacional (UPIITA-IPN), M\'exico, D.F., M\'exico
\item[$^{65}$] Universidad Aut\'onoma de Chiapas, Tuxtla Guti\'errez, Chiapas, M\'exico
\item[$^{66}$] Universidad Michoacana de San Nicol\'as de Hidalgo, Morelia, Michoac\'an, M\'exico
\item[$^{67}$] Universidad Nacional Aut\'onoma de M\'exico, M\'exico, D.F., M\'exico
\item[$^{68}$] Universidad Nacional de San Agustin de Arequipa, Facultad de Ciencias Naturales y Formales, Arequipa, Peru
\item[$^{69}$] Institute of Nuclear Physics PAN, Krakow, Poland
\item[$^{70}$] University of \L{}\'od\'z, Faculty of Astrophysics, \L{}\'od\'z, Poland
\item[$^{71}$] University of \L{}\'od\'z, Faculty of High-Energy Astrophysics,\L{}\'od\'z, Poland
\item[$^{72}$] Laborat\'orio de Instrumenta\c{c}\~ao e F\'\i{}sica Experimental de Part\'\i{}culas -- LIP and Instituto Superior T\'ecnico -- IST, Universidade de Lisboa -- UL, Lisboa, Portugal
\item[$^{73}$] ``Horia Hulubei'' National Institute for Physics and Nuclear Engineering, Bucharest-Magurele, Romania
\item[$^{74}$] Institute of Space Science, Bucharest-Magurele, Romania
\item[$^{75}$] University Politehnica of Bucharest, Bucharest, Romania
\item[$^{76}$] Center for Astrophysics and Cosmology (CAC), University of Nova Gorica, Nova Gorica, Slovenia
\item[$^{77}$] Experimental Particle Physics Department, J.\ Stefan Institute, Ljubljana, Slovenia
\item[$^{78}$] Universidad de Granada and C.A.F.P.E., Granada, Spain
\item[$^{79}$] Instituto Galego de F\'\i{}sica de Altas Enerx\'\i{}as (IGFAE), Universidade de Santiago de Compostela, Santiago de Compostela, Spain
\item[$^{80}$] IMAPP, Radboud University Nijmegen, Nijmegen, The Netherlands
\item[$^{81}$] KVI -- Center for Advanced Radiation Technology, University of Groningen, Groningen, The Netherlands
\item[$^{82}$] Nationaal Instituut voor Kernfysica en Hoge Energie Fysica (NIKHEF), Science Park, Amsterdam, The Netherlands
\item[$^{83}$] Stichting Astronomisch Onderzoek in Nederland (ASTRON), Dwingeloo, The Netherlands
\item[$^{84}$] Universiteit van Amsterdam, Faculty of Science, Amsterdam, The Netherlands
\item[$^{85}$] Case Western Reserve University, Cleveland, OH, USA
\item[$^{86}$] Colorado School of Mines, Golden, CO, USA
\item[$^{87}$] Department of Physics and Astronomy, Lehman College, City University of New York, Bronx, NY, USA
\item[$^{88}$] Louisiana State University, Baton Rouge, LA, USA
\item[$^{89}$] Michigan Technological University, Houghton, MI, USA
\item[$^{90}$] New York University, New York, NY, USA
\item[$^{91}$] Pennsylvania State University, University Park, PA, USA
\item[$^{92}$] University of Chicago, Enrico Fermi Institute, Chicago, IL, USA
\item[$^{93}$] University of Delaware, Department of Physics and Astronomy, Bartol Research Institute, Newark, DE, USA
\item[$^{94}$] University of Wisconsin-Madison, Department of Physics and WIPAC, Madison, WI, USA
\item[] -----
\item[$^{a}$] Fermi National Accelerator Laboratory, Fermilab, Batavia, IL, USA
\item[$^{b}$] Max-Planck-Institut f\"ur Radioastronomie, Bonn, Germany
\item[$^{c}$] School of Physics and Astronomy, University of Leeds, Leeds, United Kingdom
\item[$^{d}$] Colorado State University, Fort Collins, CO, USA
\item[$^{e}$] now at Hakubi Center for Advanced Research and Graduate School of Science, Kyoto University, Kyoto, Japan
\item[$^{f}$] also at University of Bucharest, Physics Department, Bucharest, Romania
\end{description}